\documentclass[preprints,accept,pdftex,moreauthors]{Definitions/mdpi} 

\firstpage{1} 

	\usepackage{upgreek}
	\usepackage{bm}
	\usepackage{braket}
	\usepackage{tikz}
	
	\newcommand*\diff{\mathop{}\!\mathrm{d}}
	\newcommand*\Diff[1]{\mathop{}\!\mathrm{d^#1}}
	
	\renewcommand{\Re}{\operatorname{Re}}
	\renewcommand{\Im}{\operatorname{Im}}
	\newcommand{\sgn}{\operatorname{sgn}}
	
	\Title{On the size of superconducting islands on the density-wave background in organic metals}
	
	\TitleCitation{On the size of superconducting islands on the density-wave background in organic metals}
	
	\Author{
		Vladislav~D.~Kochev 
		$^{1}$\orcidA{}, 
		Seidali~S.~Seidov $^{1}$\orcidC{} and 
		Pavel~D.~Grigoriev 
		$^{1,2,}$*\orcidB{}}
	
	\AuthorNames{Vladislav~D.~Kochev, Seidali~S.~Seidov, and Pavel~.D.~Grigoriev}
	
	\AuthorCitation{Kochev, V.D.; Seidov, S.S.; Grigoriev, P.D.}
	
	\address{%
		$^{1}$ \quad National University of Science and Technology ''MISiS'', 119049 Moscow, Russia; \linebreak vd.kochev@gmail.com (V.D.K.); alikseidov@yandex.ru (S.S.S.)\\
		$^{2}$ \quad L.D. Landau Institute for Theoretical Physics, 142432 Chernogolovka, Russia}

\corres{Correspondence: grigorev@itp.ac.ru}

\abstract{Most high-$T_c$ superconductors are spatially inhomogeneous. Usually, this heterogeneity originates from the interplay of various types of electronic ordering. It affects various superconducting properties, such as the transition temperature, the magnetic upper critical field, the critical current, etc. In this paper, we analyze the parameters of spatial phase segregation during the first-order transition between superconductivity (SC) and a charge- or spin-density wave state in quasi-one-dimensional metals with imperfect nesting, typical of organic superconductors. 
	An external pressure or another driving parameter increases the transfer integrals in electron dispersion, which only slightly affects SC but violates the Fermi surface nesting and suppresses the density wave (DW). At a critical pressure $P_{c}$, the transition from a DW to SC occurs. 
	We estimate the characteristic size of superconducting islands during this phase transition in organic metals in two ways. 
	Using the Ginzburg--Landau expansion, we analytically obtain a lower bound for the size of SC domains. 
	To estimate a more  specific interval of the possible size of the superconducting islands in (TMTSF)$_2$PF$_6$ samples, we perform numerical calculations of the percolation probability via SC domains and compare the results with experimental resistivity data. This helps to develop a consistent microscopic description of SC spatial heterogeneity in various \mbox{organic superconductors}.}


\begin{document}
	
	
	\section{Introduction}
	
	Superconductivity (SC) often competes~\cite{Review1Gabovich, ReviewGabovich2002,MonceauAdvPhys} with charge-density wave (CDW) or spin-density wave (SDW) electronic instabilities~\cite{Gruner1994,MonceauAdvPhys}, as~both create an energy gap on the Fermi level. 
	In such materials, the density wave (DW) is suppressed by some external parameter, which deteriorates the nesting property of the Fermi surface (FS) and enables superconductivity. 
	The driving parameters are usually the chemical composition (doping level) and pressure, as~in cuprate-~\cite{XRayNatPhys2012,XRayPRL2013,XRayCDWPRB2017,Tabis2014,Science2015Nd,Wen2019} or iron-based high-$T_c$ superconductors~\cite{ReviewFePnictidesAbrahams, ReviewFePnictides2}, organic superconductors (OSs) \cite{Ishiguro1998,AndreiLebed2008-04-23,Naito2021,Yasuzuka2009,CLAY2019,Hc2Pressure,Vuletic,Kang2010,ChaikinPRL2014, LeeBrownNMRDomains,LeeTriplet,LeeBrownNMRTriplet, CDWSC, Itoi2022}, transition metal dichalcogenides~\cite{Review1Gabovich, ReviewGabovich2002,MonceauAdvPhys}, etc. 
	The DW can also be suppressed~\cite{NbSe2NatComm} or enhanced~\cite{Gerasimenko2014, Yonezawa2018} by disorder. 
	The latter happens, e.g.,~in (TMTSF)$_{2}$ClO$_{4}$ organic superconductors~\cite{Ishiguro1998,AndreiLebed2008-04-23,Gerasimenko2013,Gerasimenko2014, Yonezawa2018}, where the disorder is controlled by the cooling rate during the anion ordering transition. 
	Anion ordering splits the electron spectrum, which deteriorates the FS nesting and dampens the SDW, enabling~SC.
	
	The SC--DW interplay is much more interesting than just a competition. 
	Usually, the~SC transition temperature, $T_{c}$, is the highest in the coexistence region near the quantum critical point where the DW disappears~\cite{Review1Gabovich, ReviewGabovich2002,Kang2010,ChaikinPRL2014}. 
	This is attributed to the enhancement of Cooper pairing by the critical DW fluctuations, similar to cuprate high-$T_c$ superconductors~\cite{Chubukov2015}. This enhancement is also common for other types of quantum critical points, such as antiferromagnetic  (AFM) points in cuprate~\cite{ArmitageReview2010,Helm2015} or heavy fermion~\cite{Mukuda2008} superconductors, ferromagnetic points~\cite{Manago2019}, nematic phase transitions in Fe-based superconductors~\cite{Eckberg2020,PhysRevX.13.011032}, etc.
	The enhancement of electron--electron ($e$--$e$) interactions in the Cooper channel already appears  in the random-phase approximation, and~the resulting strong momentum dependence of $e$--$e$ coupling may lead to unconventional superconductivity~\cite{Tanaka2004}. The~spin-dependent coupling to an SDW may additionally affect the SC in the case of their microscopic coexistence and even favor  triplet SC pairing~\cite{GGPRB2007,GrigorievPRB2008}. Generally, any antiferromagnetic background changes the spin structure of eigenstates and the electronic g-factor, as~was studied both theoretically and experimentally in cuprate and organic superconductors~\cite{Ramazashvili2021}. 
	The upper critical field $H_{c2}$ is often several times higher in the coexistence region than in a pure SC phase~\cite{Hc2Pressure,CDWSC}, which may be useful for~applications. 
	
	OSs are helpful for investigating the SC--DW interplay because they have rather weak electronic correlations and low DW and SC transition temperatures~\cite{Ishiguro1998, AndreiLebed2008-04-23}, which is convenient for their theoretical and experimental study. However, their phase diagram, layered crystal structure and many other
	features are very similar to those of high-$T_c$ superconductors. 
	Moreover, by~changing the chemical composition or pressure in OSs, one can easily vary the electronic dispersion in a wide interval and even change the FS topology from quasi-1D (Q1D) to quasi-2D. 
	Large and pure monocrystals of organic metals can be synthesized, so that their electronic structure can be experimentally studied by high-magnetic-field tools~\cite{Kartsovnik2004Nov} and by other experimental techniques~\cite{Ishiguro1998, AndreiLebed2008-04-23}.

	To understand the DW--SC interplay and the influence of DW on SC properties in OSs, one needs to know the microscopic structure of their coexistence.
	Each of these ground states creates an energy gap on the Fermi level and removes the FS instability. 
	Hence, the~DW and SC must be somehow separated in the momentum or coordinate space. 
	The momentum space DW--SC separation assumes a spatially uniform structure, where the FS is only partially gapped by the DW, and~the non-gapped parts of the FS maintain SC~\cite{MonceauAdvPhys, GrigorievPRB2008}. 
	The resistivity hysteresis observed in (TMTSF)$_{2}$PF$_{6}$ \cite{Vuletic} suggests the spatial DW/SC segregation in OSs. 
	Microscopic SC domains of size $d$  comparable to the DW coherence length $\xi_{\textrm{DW}}$ may emerge due to the soliton DW structure~\cite{BrazKirovaReview, SuReview, GrigPhysicaB2009, GG, GGPRB2007}. 
	However, such a small size of SC or metallic domains contradicts the angular magnetoresistance oscillations (AMROs) in the region of SC/DW coexistence, observed both in (TMTSF)$_{2}$ClO$_{4}$ \cite{Gerasimenko2013} and in \mbox{(TMTSF)$_{2}$PF$_{6}$ \cite{ChaikinPRL2014}} and implying the domain width $d > 1$ \textmu m~\cite{Gerasimenko2013,ChaikinPRL2014}. 
	
	The observed~\cite{Hc2Pressure,CDWSC} enhancement of the SC upper critical field $H_{c2}$ in OSs is possible in all of the above scenarios~\cite{GrigorievPRB2008, GrigPhysicaB2009}. 
	Spatial DW--SC segregation  only requires a SC domain on the order of the penetration depth $\lambda$ of the magnetic field into the superconductor~\cite{Tinkham}. 
	In (TMTSF)$_{2}$ClO$_{4}$, the penetration depth within the TMTSF layers is~\cite{Pratt2013} $\lambda_{ab} ( T=0 ) \approx 0.86$~\textmu m, and increases with $T \to T_{cSC}$. 
	Hence, the~macroscopic spatial phase separation with a SC domain size $d>1$ \textmu m suggested by AMRO data~\cite{Gerasimenko2013,ChaikinPRL2014} is consistent with the observed $H_{c2}$ enhancement in the DW--SC coexistence~phase.
	
	Another interesting feature of SDW/SC coexistence in OSs is the anisotropic SC onset, opposite to a weak intrinsic interlayer Josephson coupling in high-$T_{c}$ superconductors~\cite{Tinkham}; the SC transition and the zero resistance in OSs was first observed~\cite{Kang2010,Gerasimenko2014,ChaikinPRL2014} only along the least-conducting interlayer $z$-direction, then along the two least-conducting directions, $z$ and $y$, and~only finally in all three directions. 
	This anisotropic SC onset was explained recently~\cite{Kochev2021} by assuming a spatial SC/DW separation and studying the percolation in finite-size samples with a thin elongated shape relevant to the experiments on (TMTSF)$_{2}$PF$_{6}$ \cite{Kang2010,ChaikinPRL2014} and (TMTSF)$_{2}$ClO$_{4}$ \cite{Gerasimenko2014, Yonezawa2018}. 
	This additionally supports the scenario of spatial SC/DW segregation in the form of rather large domains of width $d>1$ \textmu m.
	However, the~microscopic reason for such phase segregation remains unknown. 
	Similar anisotropic SC onset and even $T_{c}$ enhancement in FeSe mesa structures was observed and explained by heterogeneous SC inception~\cite{Grigoriev2023FeSe}. 
	The spatial segregation in FeSe and some other Fe-based high-$T_c$ superconductors probably originates from the so-called nematic phase transition and domain structure, but~similar electronic ordering is absent in~OSs. 
	
	Recently, the~DW--metal phase transition in OSs was shown to be of first order~\cite{seidov2023firstorder}, which suggests that the spatial DW--SC segregation may be due to  phase nucleation during this transition.
	In this paper, we estimate the typical size of superconducting islands in organic metals with two different methods.  
	In Section~\ref{sec:model}, we formulate a model and the Landau--Ginzburg functional for free energy in the DW state.
	In Section~\ref{sec:sizeLandau}, we analytically obtain a lower bound for the size of the superconducting islands. 
	In Section~\ref{sec:sizeRelation}, we discuss the relationship between the DW coherence length and the SC nucleation size during the first-order phase transition.
	In Section~\ref{sec:sizePerc}, we perform numerical calculations of the percolation probability, from which we determine the interval of possible sizes of the superconducting islands in (TMTSF)$_{2}$PF$_{6}$.
	In Section~\ref{sec:discussion}, we discuss our results in connection with the experimental observations of (TMTSF)$_{2}$PF$_{6}$ and in other~superconductors.
	
	\section{The~Model}\label{sec:model}
	\subsection{Q1D Electron Dispersion and the Driving Parameters of DW--Metal/SC Phase Transitions in~OSs}
	
	In Q1D organic metals~\cite{Ishiguro1998, AndreiLebed2008-04-23}, the free electron dispersion near the Fermi level 
	is approximately given by 
	\begin{equation}
		\varepsilon (\bm{k})=\hbar v_{F}(|k_{x}|-k_{F})
		+ t_{\perp}(\bm{k}_{\perp }),  \label{dispersion1}
	\end{equation}
	where $v_{F}$ and $k_{F}$ are the Fermi velocity and Fermi momentum in the chain $x$-direction. 
	The interchain electron dispersion $t_{\perp }(\bm{k}_{\perp })$ is given by the tight-binding model:
	\begin{equation}
		t_{\perp }(\bm{k}_\perp)= 
		2 t_{b}\cos (k_y b)+2 t_b^{\prime} \cos (2k_y b),  \label{dispersion}
	\end{equation}%
	where $b$ is the lattice constant in the $y$-direction. 
	The dispersion along the interlayer $z$-axis is usually significantly less than along the $y$-axis; thus, it is left out here. 
	In (TMTSF)$_{2}$PF$_{6}$, the transfer integral $t_{b}$ is $\approx 30$~meV~\cite{Valfells1996}, and~the ''antinesting'' parameter $t_b'$ is $\approx 4.5$~K~\cite{Chaikin1996} at ambient~pressure.
	
\begin{figure}[H]
		\centering
		\begin{tikzpicture}[every node/.style={inner sep=0,outer sep=0}]
			\node (picture) {\includegraphics{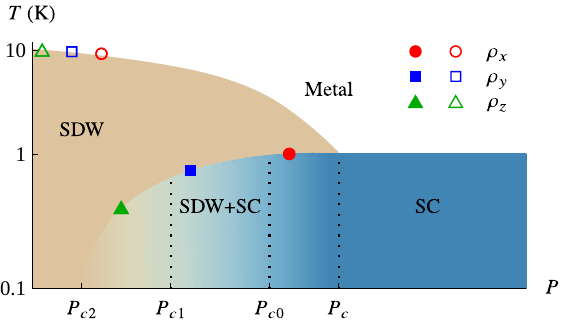}};
			\node[below=0.2cm,left=0.5cm] at (picture.north east) {(a)};
		\end{tikzpicture}
		\quad
		\begin{tikzpicture}[every node/.style={inner sep=0,outer sep=0}]
			\node (picture) {\includegraphics{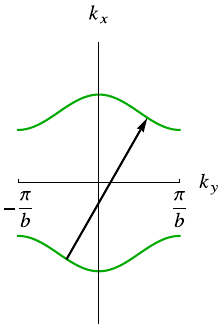}};
			\node[below=0.2cm,left=0.5cm] at (picture.north east) {(b)};
		\end{tikzpicture}
		
		\caption{(\textbf{a}) Pressure--temperature phase diagram of
			(TMTSF)$_{2}$PF$_{6}$ recreated from resistivity data in ref.~\cite{Kang2010};
			(\textbf{b}) schematic FS of (TMTSF)$_{2}$PF$_{6}$, obtained from the Q1D electron dispersion given by Equations~(\ref{dispersion1}) and (\ref{dispersion}). The~nesting vector $\bm{Q}$ is indicated by the black arrow.
		}
		\label{FigPhaseDiagramAndFS}
	\end{figure}

	As illustrated in Figure~\ref{FigPhaseDiagramAndFS}b, the~FS of Q1D metals consists of two slightly warped sheets separated by $2 k_F$ and roughly exhibits the nesting~property.
	It leads to the Peierls instability and favors the formation of DWs at low temperatures $T < T_{\textrm{cDW}}\equiv T_{\textrm{c}}$, which competes with superconductivity. 
	The quasiparticle dispersion in the DW state in the mean-field approximation is given by
	\begin{equation}
		E_\pm (\bm{k}) = \varepsilon_+ (\bm{k},\bm{k}-\bm{Q}) 
		\pm \sqrt{|\Delta_{\bm{Q}}|^2 + \varepsilon_-^2 (\bm{k},\bm{k}-\bm{Q})},  \label{EnSp}
	\end{equation}%
	where we have used the notations
	\begin{equation}
		\varepsilon_\pm (\bm{k},\bm{k}^{\prime })=
		\frac{\varepsilon ({\bm{k}})\pm \varepsilon ({\bm{k}}^{\prime }) }
		{2}.  \label{epspm}
	\end{equation}%
	
	The FS has the property of perfect nesting at the wave vector $\bm{Q}$ if $\varepsilon _{+}(\bm{k},\bm{k}-\bm{Q}) =0$. 
	If $\varepsilon_{+}({\bm{k}},{\bm{k}}-{\bm{Q}})< |\Delta_{\bm{Q}}|$ for the entire FS, all electron states are gapped at the Fermi level due to  DW formation. 
	Then, the DW converts to a semiconducting state at $T<T_{\textrm{cDW}}$ and SC does not emerge. 
	If $\varepsilon_{+}({\bm{k}},{\bm{k}}-{\bm{Q}})>|\Delta _{{\bm{Q}}}|$ in a finite interval of ${\bm{k}}$ at the Fermi level, the~metallic state survives at $T<T_{\textrm{cDW}}$. 
	Then, a uniform SC state may emerge, but~its properties differ from those without DWs~\cite{GGPRB2007,GrigorievPRB2008} because of the FS reconstruction and the change in electron dispersion by the DW. 
	For ${\bm{Q}}={\bm{Q}}_{0}=\left( 2k_{F},\pi /b\right) $, only the second harmonic in the electron dispersion given by Equation~(\ref{dispersion}) violates  FS nesting: $\varepsilon _{+}({\bm{k}},{\bm{k}}-{\bm{Q}}_{0})= 2t_{b}^{\prime }\cos (2k_{y}b)$. 
	Hence, usually only $t_{b}^{\prime} \sim t_{b}^{2}/v_{F}k_{F}\ll t_{b}$ is important for the DW phase~diagram.
	
	With the increase in applied pressure $P$, the lattice constants decrease.
	This enhances the interchain electron tunneling and the transfer integrals.
	The increase in $t_{b}^{\prime } (P) $ with pressure spoils the FS nesting and decreases the DW transition temperature $T_{\textrm{cDW}} (P)$. 
	There is a critical pressure $P_{c}$ and a corresponding critical value $t_{b}^{\prime \ast }=t_{b}^{\prime } (P_{c}) $ at which $T_{\textrm{cDW}}\left( P_{c}\right) =0$ and a quantum critical point (QCP) exists. 
	The electronic properties at this DW QCP are additionally complicated by superconductivity emerging at $T<T_{\textrm{cSC}}$ at $P>P_{c}$. 
	In organic metals, SC appears even earlier, at~$P>P_{c1}<P_{c}$, and~there is a finite region $P_{c1}<P<P_{c}$ of SC--DW coexistence~\cite{Kang2010, ChaikinPRL2014,CDWSC}.
	This simple model qualitatively describes the phase diagram observed in (TMTSF)$_{2}$PF$_{6}$ \cite{Kang2010, ChaikinPRL2014,CDWSC}, $\upalpha$-(BEDT-TTF)$_{2}$KHg(SCN)$_{4}$ \cite{CDWSC}, in~various compounds of the (TMTTF)$_{2}$X family~\cite{Itoi2022,Araki2007,Auban2003} and in many other OSs~\cite{Ishiguro1998, AndreiLebed2008-04-23,Yasuzuka2009,CLAY2019}.
	
	\subsection{Mean Field Approach and the Landau--Ginzburg Expansion of DW
		Free~Energy}\label{sec:meanfield}
	
	Mean-field theory does not  correctly describe strictly 1D
	conductors, where  non-perturbative methods are helpful 
	. However, in~most DW materials,  nonzero
	electron hopping between the conducting 1D chains and the 3D character of
	the electron--electron (e--e) interactions and lattice elasticity reduce the
	deviations from the mean-field solution and also make most of the methods
	and exactly solvable models developed for the strictly 1D case inapplicable. On~the other hand, the~interchain electron dispersion strongly dampens the
	fluctuations and validates the mean-field description \cite%
	{Horovitz1975,McKenzie1995}. The~perpendicular-to-chain term $t_{\perp }(%
	\bm{{k}_{\perp })}$ in Equations~(\ref{dispersion1}) and (\ref{dispersion})
	is much greater than the energy scale of the DW transition temperature ($%
	T_{c0} \approx 12.1$~K). Only the \textquotedblright imperfect nesting\textquotedblright\
	term $\sim t_{b}^{\prime }$ of $t_{\perp }(\bm{{k}_{\perp })}$ is on the
	order of $T_{c0}$
	. Hence, the~criterion for the mean-field theory to be applicable~\cite{Horovitz1975,McKenzie1995}, $t_{\perp }\gg T_{c0}$, is reliably satisfied in
	most Q1D organic~metals.
	
	For our analysis, we take the Landau--Ginzburg expansion of the free energy in the
	series of even powers of the DW order parameter $\Delta = \Delta_{\bm{Q}}$:
	\begin{equation}
		F \simeq \frac{A (T,\bm{Q})}{2} |\Delta|^2 +\frac{B}{4} |\Delta|^{4}
		+\frac{C}{6} |\Delta|^{6} +\frac{D}{8} |\Delta|^{8} +\dots
		\label{F}
	\end{equation}
	
	Usually, the~minimum of the free energy corresponds to the uniform DW order parameter $\Delta $ when ${\bm{Q=Q}}_{0}$. 
	Since the coefficient $A\left( T_\text{cDW},{\bm{Q}}_{0}\right) =0$, we keep its temperature and momentum dependence. 
	The sign of the coefficient $B$ determines the type of DW--metal phase transition. 
	If $B_\text{DW}>0$, the~phase transition is of the second order, and~only the first two coefficients $A_\text{DW}$ and $B_\text{DW}$ are sufficient for its description. 
	If $B_\text{DW}<0$, the~phase transition may be of the first order and the coefficients $C_\text{DW}$ and even $D_\text{DW}$ if $C_\text{DW}\leq 0$ are required for its description. 
	The self-consistency equation (SCE) for a DW is obtained by the variation in the free energy (\ref{F}) with respect to $\Delta$:
	\begin{equation}
		\Delta \left( A +B\left\vert \Delta
		\right\vert ^{2}+C\left\vert \Delta \right\vert ^{4}+D\left\vert \Delta
		\right\vert ^{6}+\dots \right) =0.  \label{SC}
	\end{equation}%
	
	The free energy (\ref{F}) can also be calculated by integrating the SCE over $\Delta$. In~ref.~\cite{Grigoriev2005}, the SCE for the DW was derived in a magnetic field acting via Zeeman splitting and for two coupling constants of the $e$--$e$ interaction, charge $U_{c}$ and spin $U_{s}$ (see Equations~(17) in ref.~\cite{Grigoriev2005}). 
	Without a magnetic field, the charge $U_{c}$ and spin $U_{s}$ coupling constants do not couple, and~the system chooses the largest one of them, corresponding to the highest transition temperature. 
	We rewrite the SCE without a magnetic field and for only one charge or spin coupling constant $U$:%
	\begin{equation}
		\Delta =-TU \sum_{\bm{k}\omega}
		\frac{\Delta}{(\omega +i \varepsilon
			_{+})^2 +\varepsilon_{-}^2 +|\Delta|^2},  \label{Delta_SC}
	\end{equation}%
	where $\varepsilon_\pm = \varepsilon_\pm (\bm{k},\bm{k}-\bm{Q})$ are given by Equation~(\ref{epspm}), and~$\omega$ takes the values $\pi T(2n+1)$, $n\in \mathbb{Z}$. 
	In Appendix \ref{sec:AppendixA}, we briefly describe the derivation of Equation~(\ref{Delta_SC}) and discuss the relation of coefficients in the Landau--Ginzburg expansion (\ref{F}) with electronic susceptibility.
	The Landau--Ginzburg expansion coefficients in Equations~(\ref{F}) and (\ref{SC}) can be obtained by the expansion of Equation~(\ref{Delta_SC}) in a power series of $|\Delta |^{2}$. 
	
	The sum over $\bm{k}$ in Equation~(\ref{Delta_SC}) for a macroscopic sample is equivalent to the integral:
	\begin{equation}
		\sum_{\bm{k}}=
		2 \int \frac{\diff k_x}{2\pi} \int_{-\pi/b}^{\pi/b} \frac{\diff k_y}{2\pi}.
	\end{equation}
	
	The factor of $2$ appears because of two FS sheets are present at $k_{x} \approx \pm k_{F}$. Usually, for~simplicity, the integration limits over $k_{x}$ are taken to be infinite and the resulting logarithmic divergence of Equation~(\ref{Delta_SC}) is regularized by the definition of the transition temperature $T_{c0}$. This procedure is briefly described in Appendix B of ref.~\cite{seidov2023firstorder}. When the Fermi energy $E_{F}\gg t_{b}$, for~a linearized electron dispersion (\ref{dispersion1}) near the Fermi level, one may integrate Equation~(\ref{Delta_SC}) over $k_{x}$ in infinite limits, which gives (cf. Equation~(22) of ref.~\cite{Grigoriev2005})
	\begin{equation}
		\Delta =\frac{\pi \nu_{F} |U| T}{2} \sum_{\omega} \left\langle \frac{\Delta}
		{\sqrt{ (\omega +i \varepsilon_+)^2 +|\Delta|^2}} \right\rangle
		_{k_{y}},  \label{SC0}
	\end{equation}%
	where the density of electron states at the Fermi level in the metallic phase per two spin components per unit length $L_{x}$ of one chain is $\nu _{F}=2/\pi \hbar v_{F}$. Averaging over $k_{y}$ is denoted by triangular brackets, i.e.,~$\braket{ \cdot }_{k_y} = b \int_{-\pi /b}^{\pi /b} \diff k_y /2\pi \cdot$. 
	Equation~(\ref{SC0}) is similar to the self-consistency equation for superconductivity in a magnetic field, where the orbital effect of the magnetic field is neglected and the pair-breaking Zeeman splitting is \mbox{replaced by $\varepsilon_{+}({\bm{k}},{\bm{k}}-{\bm{Q}})$}.

	\section{Estimation of the Size of the SC~Islands}\label{sec:size}
	
	\subsection{Analytical Calculation of the Ginzburg--Landau Expansion Coefficients for $T \gg {t_b}'$}\label{sec:sizeLandau}
	
	Expansion of Equation~(\ref{SC0}) over $\Delta $ yields%
	\begin{equation}
		\begin{split}
			A &= - \frac{4 \pi T}{\hbar v_{F}} \sum_{\omega } \left\langle \frac{\sgn \omega }{
				\omega +i \varepsilon_+}\right\rangle_{k_y} -\frac{1}{U}= \\
			& = - \frac{4}{\hbar v_{F}}\left[ \ln \frac{T_{c0}}{T}
			+\psi \left( \frac{1}{2}\right)
			-\left\langle\Re \psi \left( \frac{1}{2}+\frac{\varepsilon_+}{2\pi T}%
			\right) \right\rangle _{k_{y}}  \right], \label{A}
		\end{split}
	\end{equation}
	where the logarithmic divergence $\pi T\sum_{\omega}\left\vert \omega \right\vert ^{-1}\approx \ln \left( E_{F}/T\right) $ is contained in the definition of $T_{c0}=E_{F}\exp \left\{ -1/\left( v_{F}|U|\right) \right\}$. In~(TMTSF)$_2$PF$_6$, $T_{c0}\approx 12.1$~K~\cite{Chaikin1996}.

	The spatial modulation with the wave vector $\bm{q}$ of the DW order parameter $\Delta$ corresponds to the deviation of the DW wave vector $\bm{Q}$ from $\bm{Q}_0$ by $\pm \bm{q}$. Hence, the~gradient term in the Ginzburg--Landau expansion of the  DW free energy can be obtained by the expansion of $A(T,\bm{Q})$ given by Equation~(\ref{A}) in the powers of small deviation $\bm{q}=\bm{Q}-\bm{Q}_0$. 
	$A(\bm{Q},T)$ depends on $\bm{Q}$ via $\varepsilon_+ =\varepsilon_+ (\bm{k},\bm{k}-\bm{Q}) $, given by Equation~(\ref{epspm}). 
	For the quasi-1D electron dispersion in Equations~(\ref{dispersion1}) and (\ref{dispersion}), approximately describing (TMTSF)$_2$PF$_6$, we may use Equation~(21) from ref.~\cite{Grigoriev2005}:
	\begin{equation}
		\varepsilon_+ = \frac{h v_F q_x}{2}+ 2 t_b \sin \frac{b q_y}{2} \sin \left(b \left[k_y-\frac{q_y}{2}\right]\right)
		- 2t_b'\cos \left(b q_y\right)\cos \left(b \left[2 k_y-q_y\right]\right) . \label{ePlus}
	\end{equation}
	
	The general form of the Taylor series of $A(T,\bm{Q}_0+\bm{q})$, given by Equation~(\ref{A}), over~the deviation $\bm{q=Q-Q}_{0}$ of the DW wave vector $\bm{Q}$ from its optimal value $\bm{Q}_{0}$ up to the second order is
	\begin{equation}
		A(\bm{q}) \simeq - \frac{4}{\hbar v_F} \left[ \ln \frac{T_{c0}}{T} 
		+ \int\limits_{-\pi/b}^{\pi/b} \diff k_y 
		\left( C_0 + c_x q_x + c_y q_y + c_{xy} q_x q_y + A_x q_x^2 + A_y q_x^2 \right)
		\right]. 
	\end{equation}
	
	The linear terms and the cross term vanish when taking the integral $\int_{-\pi/b}^{\pi/b} \diff q_y$ (this is always the case if wave vector $\bm{Q}_0$ is the optimal one). The~constant and quadratic terms do not vanish, thus
	\begin{equation}	
		A(\bm{q}) \simeq A_0 +  A_x q_x^2 +  A_y q_y^2.
	\end{equation}
	
	Expanding Equation~(\ref{ePlus}) over the deviation $\bm{q=Q-Q}_{0}$ up to the second order, substituting it in Equation~(\ref{A}) and expanding the digamma function over the same wave vector $\bm{q}$, we obtain the coefficients $A_i$:
	\begin{equation}
		\begin{split}
			A_0 &= -\frac{4}{\hbar v_F} 
			\left[ \frac{T_{c0}}{T} + \psi \left(\frac{1}{2}\right) 
			- \Braket{ \Re \psi \left(\frac{1}{2}-\frac{i t_b' \cos \left(2 b k_y\right)}{\pi T}\right)
			}_{k_y} \right]; \\
			A_x &= -\frac{4}{\hbar v_F} 
			\frac{\hbar^2 v_F^2}{32 \pi^2 T^2}   \Braket{ 
				\Re \psi^{(2)} \left(\frac{1}{2}-\frac{i t_b' \cos \left(2 b k_y\right)}{\pi T}\right)
			}_{k_y} ; \\
			A_y &= \frac{4}{\hbar v_F} 
			\frac{b^2}{8 \pi^2 T^2}   \Bigg\langle  
			2 \pi  T \left[t_b \cos \left(b k_y\right)-4 t_b' \cos \left(2 b k_y\right)\right] \Im \psi^{(1)} \left(\frac{1}{2}-\frac{i t_b' \cos \left(2 b k_y\right)}{\pi  T}\right) -\\
			&-\left[t_b-4 t_b' \cos \left(b k_y\right)\right]^2 \sin^2\left(b k_y\right) \Re\psi^{(2)}\left(\frac{1}{2}-\frac{i t_b' \cos \left(2 b k_y\right)}{\pi  T}\right) 
			\Bigg\rangle_{k_y}.
		\end{split} \label{Ai_num}
	\end{equation}
	
	The integrals over $k_y$ in $A_y$, $A_x$ and $A_0$ can be calculated numerically and give the coherence lengthes $\xi_x$ and $\xi_y$. From~Equations~(\ref{chi1}) and (\ref{chiQ}), it follows that
	\begin{equation}
		\xi^2_i = A_i/A_0. \label{xi}
	\end{equation} 
	
	Figure~\ref{FigXi} shows $\xi_x$ and $\xi_y$ as functions of temperature $T$ for two different values of $t_b'$: $t_b'\approx 0.42~t_b^{'\ast}$, corresponding to (TMTSF)$_2$PF$_6$ at ambient pressure~\cite{Chaikin1996} (solid orange and dashed green lines), and~$t_b'=0.95~t_b^{'\ast}$ (solid red and dashed blue lines), i.e.,~close to the quantum critical point at $t_b'=t_b^{'\ast}$. These curves diverge at $T=T_c(t_b')$, where $A_0=0$. This divergence, being a general property of phase transitions, is well known in superconductors. When plotting Figure~\ref{FigXi}, we used Equations~(\ref{Ai_num}) and (\ref{xi}), and~the parameters of (TMTSF)$_2$PF$_6$, i.e., $b = 0.767$~nm~\cite{Kim_2009} and $v_F = 10^7$~cm/s~\cite{Valfells1996}.
	

	At rather high temperatures ($2\pi T \gg \varepsilon_+ \sim {t_b}'$), we may expand the digamma function in Equation~(\ref{A}) to a Taylor series near  $1/2$, which gives
	\begin{equation}
		\psi\left(\frac{1}{2}\right) - \Braket{\Re \psi \left(
			\frac{1}{2} + \frac{i \varepsilon_+ (\bm{k},\bm{k}-\bm{Q})}{2 \pi T}
			\right)}_{k_y} \simeq 
		\frac{b}{2\pi} \int\limits_{-\pi/b}^{\pi/b} \!\diff k_y \frac{\psi^{(2)}(1/2)  \varepsilon_+^2 (\bm{k},\bm{k}-\bm{Q})}{8 \pi^2 T^2}. \label{A2}
	\end{equation}
	
	Expanding Equation~(\ref{ePlus}) over $\bm{q}$ up to the second order and substituting it into Equation~(\ref{A}) after using Equation~(\ref{A2}), we obtain the coefficients $A_i$:
	\begin{equation}
		\begin{split}
			A_0 &= - \frac{4}{\hbar v_F} \left[ \ln \frac{T_{c0}}{T} + \frac{{t_b'}^2}{4 \pi ^2 T^2} \right]; \\
			A_x &= - \frac{4}{\hbar v_F} \psi^{(2)}\!\left(\frac{1}{2}\right) \left[ \frac{h^2 v_F^2}{32 \pi ^2 T^2} \right]; \quad
			A_y = - \frac{4}{\hbar v_F} \psi^{(2)}\!\left(\frac{1}{2}\right) \left[ \frac{b^2 \left(t_b^2-4 {t_b'}^2\right)}{16 \pi ^2 T^2} \right].
		\end{split} \label{Ai_approx}
	\end{equation}
	
	Substituting them into Equation~(\ref{xi}), we derive simple analytical formulas for the SDW coherence lengths, $\xi_x$ and $\xi_y$, valid at $\pi T\gg t_b'$:
	\begin{equation}
		\label{xi_approx}
		\begin{split}
			\xi_x &= \frac{\hbar v_F}{2 \sqrt{2}} \sqrt{\left. 1 \middle/ \left(
				4 \pi^2 T^2 \ln (T_{c0}/T) / \psi^{(2)}(1/2) + {t_b'}^2 
				\right) \right.}; \\
			\xi_y &= \frac{b}{2} \sqrt{
				\frac
				{t_b^2 - 4 {t_b'}^2}
				{4 \pi^2 T^2 \ln (T_{c0}/T) / \psi^{(2)}(1/2) + {t_b'}^2}
			}. 
		\end{split} 
	\end{equation}
	
	At this limit of $\pi T\gg t_b'$, the ratio of coherent lengths along the $y$- and $x$-axes does not depend on temperature:
	\begin{equation}
		\frac{\xi_y}{\xi_x} = \frac{b}{\hbar v_F} \sqrt{2 \left(t_b^2-4 {t_b'}^2\right)} \approx 0.5.
	\end{equation}
	
	\textls[-15]{The temperature dependence of the coherence lengthes $\xi_x$ and $\xi_y$ given by \mbox{Equation~(\ref{xi_approx})} are shown in Figure~\ref{FigXi} by dotted lines.} The~black dotted curves in Figure~\ref{FigXi} are obtained from Equation~(\ref{xi_approx}) by setting 
	$t_b'=4.5~\text{K}=0.42~t_b^{'\ast}$, corresponding to (TMTSF)$_2$PF$_6$ at ambient pressure~\cite{Chaikin1996}. These curves coincide with the result of numerical integration in Equations~(\ref{Ai_num}), which confirms the applicability of Equations~(\ref{Ai_approx}) and (\ref{xi_approx}) with these parameters. From~Equation~(\ref{xi_approx}), we obtain $\xi_x \approx 0.06$~\textmu m and $\xi_y \approx 0.03$~\textmu m at $T = T_{c0} = 12.1$~K, corresponding to $T/T_c-1\approx 0.075$. However, at~$T/T_c-1\approx 0.01$, this gives $\xi_x \approx 0.16$~\textmu m and $\xi_y \approx 0.08$~\textmu m.
	
	\subsection{Relation between the Coherence Length and Nucleation Size during the First-Order Phase~Transition}\label{sec:sizeRelation}
	
	Despite an extensive study of the phase nucleation process during the first-order phase transition~\cite{Oxtoby1998,Umantsev2012,Kalikmanov2012Nucleation,Karthika2016}, its general quantitative description is still missing. The~nucleation rate and size may strongly depend on minor factors relevant to a particular system. The~DW--metal or DW--SC phase transitions also have peculiarities, such as a strong dependence on the details of electron dispersion. Nevertheless, one can roughly estimate the lower limit of the nucleus size using the Ginzburg--Landau expansion for DW free energy. The~latter gives the energy of a phase nucleus $\Omega $, described by the spatial variation $\Delta (\bm{r})$ of the DW order parameter during the first-order phase transition as
	\begin{equation}
		\Delta F_{\Omega} \approx \int_{\Omega} \Diff{3}{\bm{r}} \frac{1}{2}\left[ A_0\Delta ^2 + \sum_i A_i (\partial_i \Delta )^2 \right] \approx \int_{\Omega} \Diff{3}{\bm{r}} \frac{A_0}{2} \left[ \Delta ^2 + \sum_i (\xi_i \partial_i \Delta )^2\right] .
	\end{equation} 
	
\begin{figure}[H]
			\centering
		\includegraphics{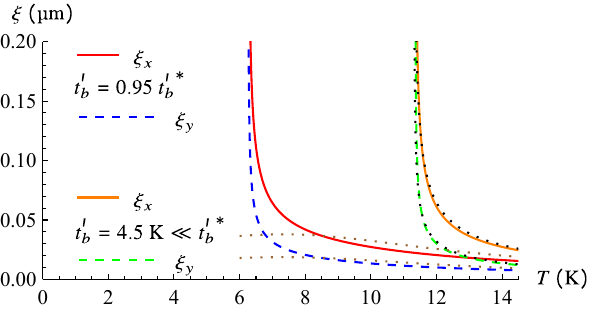}
		\caption{Temperature dependence of the DW coherence length $\xi$ along two main axes at two different values of $t_b'$: $t_b'=4.5~\text{K}=0.42~ t_b^{'\ast}$, corresponding to (TMTSF)$_2$PF$_6$ at ambient pressure, and~$t_b'=0.95~t_b^{'\ast}$. Solid and dashed lines correspond to the numerical solution of Equations (\ref{Ai_num}), while the dotted lines correspond to the approximate analytical formulas in Equations~(\ref{Ai_approx}) and (\ref{xi_approx}).}
		\label{FigXi}
	\end{figure}

	If the nucleus size $d_i$ is $<2\xi_i$, the~second (always positive) gradient term exceeds the first term, which is energetically unfavorable. Hence, the~minimal dimensions of phase nucleation during the first-order phase transition is given by the coherence lengths $d_i>2\xi_i$. The~latter diverges at the spinodal line $T_c(t'_b)$ of the phase transition where $A_0=0$, as~illustrated in Figure~\ref{FigXi} for our DW system. However, the~first-order phase transition starts at a slightly different temperature $T_{c1}$, while the spinodal line $T_c(t'_b)$ corresponds to the instability of one phase. Hence, for~the estimates of nucleus size $d_i$, one should take some finite interval $\Delta T = T_{c1}-T_c$, which is determined by the width of the first-order phase transition. Unfortunately, the~latter is unknown and strongly depends on the physical system. In~our case, this width $\Delta T$ depends on the details of electron dispersion, e.g.,~on the amplitude of higher harmonics in the electron dispersion given by Equation~(\ref{dispersion}). If~we take a reasonable estimate, i.e., $\Delta T = T_{c1}-T_c\approx 0.01~T_c$, we obtain the SC domain size \mbox{$d>2\xi >0.3$~\textmu m}. 
	
	\subsection{Estimates of Superconducting Island Size from Transport Measurements and the Numerical Calculation of the Current Percolation~Threshold}\label{sec:sizePerc}

	Another method of estimating the average SC island size is based on using the available transport measurements, especially the anisotropy of the SC transition temperature observed in various organic superconductors~\cite{Kang2010,ChaikinPRL2014,Gerasimenko2014} and determined from the anisotropic zero-resistance onset in various samples. This anisotropy was explained both in organic superconductors~\cite{Kochev2021} and in mesa structures of FeSe~\cite{Grigoriev2023FeSe} by the direct calculation of the percolation threshold along different axes in samples of various spatial dimensions relevant to experiments. The~qualitative idea behind this anisotropy is very simple. As~the volume fraction $\phi$  of the SC phase grows, the~isolated clusters of superconducting islands grow and become comparable to the sample size. When the percolation via superconducting islands between the opposite sample boundaries is established,~zero resistance sets in. If~the sample shape is flat or needle-like, as~in organic metals, this percolation first establishes along the shortest sample dimension, when the SC cluster becomes comparable to the sample thickness (see Figure~4a in ref.~\cite{Kochev2021} or Figure~4b in ref.~\cite{Grigoriev2023FeSe} for illustration). With~a further increase in the SC volume fraction $\phi$, the zero resistance sets in along two axes, and~only finally in all three  directions, including the sample~length. 
	
	In  infinitely large samples, the percolation threshold is isotropic~\cite{PercolationEfros}. Hence, this anisotropy depends on the ratio of the average size $d$ of superconducting islands to the sample size $L$. This dependence can be used for a qualitative estimate of SC island size $d$ by analyzing the interval of $d$ where the experimental data on conductivity anisotropy are consistent with theoretical~calculations.

	The algorithm and implementation details of percolation calculations are given in refs.~\cite{Kochev2021,Grigoriev2023FeSe}.
	Using this method, we calculated the probability of percolation of a random geometric configuration of superconducting islands in a sample of (TMTSF)$_2$PF$_6$ with typical experimental dimensions of $3 \times 0.2 \times 0.1$~mm$^3$ \cite{Vuletic,Kang2010} for various island sizes. For~simplicity, the~geometry of the islands was taken~as spherical. 
	
\begin{figure}[H]
			\centering
		\includegraphics{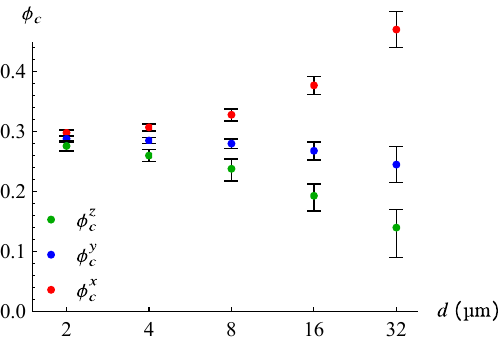}
		\caption{The dependence of the percolation threshold $\phi_c$ along different axes on the size of the spherical island $d$ in (TMTSF)$_2$PF$_6$. The~intervals of $\phi_c$, corresponding to the percolation probability $p \in (0.5 \pm 0.4)$, are indicated by error~bars. }
		\label{FigPerc}
	\end{figure}

	Figure~\ref{FigPerc} shows the dependence of the percolation threshold $\phi_c$ of the SC phase on the geometric dimensions of the superconducting islands. By~the percolation threshold, we mean the SC volume fraction $\phi$ at which the probability of percolation of a randomly chosen geometrical configuration of islands is $1/2$. In~order to take into account possible random fluctuations of this SC current percolation, in~Figure~\ref{FigPerc}, we also plot the interval of the SC volume fraction $\phi$, corresponding to the large interval of percolation probability $p\in (0.1,0.9)$ and denoted by the error bars. These error bars get bigger with the increase in size $d$ of the spherical islands, because~the larger the SC domain size $d$, the~smaller the number $N$ of SC domains  required for percolation and hence, the~stronger  its relative fluctuations $\delta N/N\propto N^{-1/2}$. From~Figure~\ref{FigPerc}, we can see that for the sample dimensions used in the experiment~\cite{Kang2010}, the~percolation threshold via SC domains is considerably anisotropic, beyond~the random fluctuations corresponding to a particular sample realization, if~the domain size exceeds 2~\textmu m. For~smaller sizes of superconducting islands, the anisotropy is smaller than the ''error bar'' of $\phi_c$, corresponding to the fluctuations in the percolation probability $p\in (0.1,0.9)$. These error bars get bigger with the increase in the size $d$ of  the superconducting islands, because~the larger the SC domain size, the~smaller the number $N$ of SC domains  required for percolation and the stronger  the fluctuations in this number $\delta N/N\propto N^{-1/2}$. For~$d< 2$~\textmu m, the percolation thresholds along all three axes converge to the known isotropic percolation threshold in infinite samples $\phi_{c\infty}\approx 0.2895$ (see page 253 of ref.~\cite{Torquato2002}). 
	\section{Discussion and~Conclusions}\label{sec:discussion}
	
	\textls[-15]{The observed strong anisotropy of the SC transition temperature \mbox{$T_\text{cSC}$ in (TMTSF)$_2$PF$_6$  \cite{Kang2010,ChaikinPRL2014}}} and (TMTSF)$_2$ClO$_4$ \cite{Gerasimenko2014} samples of thicknesses of $\sim 0.1$~mm is consistent with our percolation calculations of the SC domain size of $d>2$~\textmu m. 
	These estimates of the SC domain size $d$ agree well with the result that $d_x>1$~\textmu m, implied by the clear observation of angular magnetoresistance oscillations and of a field-induced SDW in (TMTSF)$_2$PF$_6$ \cite{ChaikinPRL2014} and  \mbox{(TMTSF)$_2$ClO$_4$ \cite{Gerasimenko2013}}. 
	The latter requires that the electron mean free path, $l_{\tau }$, is $>l_B$, where $l_B=\hbar /eBb \sim 1$~\textmu m is the so-called quasi-1D magnetic length~\cite{Kartsovnik2004Nov,ChaikinPRL2014}. 
	Hence, all experimental observations agree and suggest an almost macroscopic spatial separation of SC and SDW phases in these organic~superconductors. 
	
	The above SDW coherence length $\xi$ obtained from the Ginzburg--Landau expansion of the SDW free energy at the first-order SDW--SC phase transition in the organic superconductor (TMTSF)$_2$PF$_6$ gives the SC domain size $d>2\xi > 0.3$~\textmu m. 
	This generally agrees with the experimental estimates of $d>1$~\textmu m, but~gives a too weak limitation because of the following three possible reasons:
	\begin{enumerate}[leftmargin=*,labelsep=4.9mm]
		\item [(1)]The SC proximity effect~\cite{Tinkham}: The SC order parameter is nonzero not only in the SC domains themselves, but~also in  shells of width $\delta d\sim \xi_{SC}$ around these SC domains. 
		The SC coherence length $\xi_{SC}\sim \hbar v_F/\pi\Delta_\text{SC}$ diverges near the SC transition temperature $T_
		\text{cSC}$, and~even far from $T_\text{cSC}\approx 1$~K in organic superconductors $\delta d_x\sim \xi_{SC}\sim \hbar v_F/\pi T_{cSC}\approx 0.3$~\textmu m. 
		Hence, the~resulting size of SC domains with this proximity effect shell is $d_x \gtrsim 2(\xi_{SC}+\xi )\approx 1$~\textmu m, which well agrees with \mbox{experimental~data}.
		
		\item [(2)]The clusterization of superconducting islands with the formation of larger SC domains, glued by the Josephson junction: In~current percolation and zero-frequency transport measurements, such a cluster is seen as a single SC domain. 
		Since the SDW--SC transition is observed close to the  SC percolation threshold (the~SC volume fraction of $\phi_c>0.1$),~the formation of such SC clusters is very probable. 
		Note that such clusterization may also explain the small difference between the estimates of the SC domain size from AMRO data and from current~percolation.
		
		\item [(3)]An oversimplified physical model: In~our percolation calculations, we take all clusters of the same size, because~the actual size distribution of superconducting islands is unknown. 
		In addition, special types of disorder, such as local variations in (chemical) pressure, affect the SC--SDW balance.  
	\end{enumerate}
	
	
	The size of isolated SC domains allows an independent approximate measurement of the diamagnetic response. 
	The diamagnetic response of small SC grains of size $d\lesssim \lambda$, where $\lambda$ is the penetration depth of magnetic field into the superconductor, strongly depends on the  $d/\lambda$ \cite{Tinkham} ratio. Note that this penetration depth in layered superconductors is anisotropic. 
	Since the SC volume fraction $\phi$ is approximately known from the transport measurements and from the percolation threshold, by~measuring the diamagnetic response at $\phi <\phi_c$ for three main orientations of a magnetic field $\bm{B}$ and comparing it with the susceptibility $\chi =-\phi /(4\pi )$ of large SC domains of volume fraction $\phi$, one may roughly estimate the SC domain size along all three axes. 
	A similar diamagnetic response in combination with transport measurements was used in FeSe to estimate the size and shape of superconducting islands above $T_\text{cSC}$ \cite{Sinchenko2017,Grigoriev2017}. 
	A similar combined analysis of the diamagnetic response and transport measurements has also been used to obtain information about the SC domain size and shape above $T_\text{cSC}$ in another organic superconductor, $\beta $-(BEDT-TTF)$_{2}$I$_{3}$ \cite{Seidov2018}. 
	
	Spatial phase segregation may also happen near the quantum critical point of the Mott-AFM metal--insulator phase transition, e.g.,~as observed in the $\kappa$-(BEDT-TTF)$_2$X family of organic superconductors~\cite{Miyagawa2002,Sasaki2009,Zverev2019,Kartsovnik2023}.  The~observation of clear magnetic quantum oscillations~\cite{Zverev2019,Kartsovnik2023} in the almost insulating phase of these materials indicates a rather large size $d$ of metal/SC domains in the Mott insulator media, comparable to the electron cyclotron radius. Although~the first of our methods, based on the Ginzburg--Landau SDW free energy expansion, is not applicable in this case, our second method~\cite{Kochev2021,Grigoriev2023FeSe}, based on the calculation of percolation anisotropy in finite-sized samples, should work well and give  valuable information about the shape and size of metal/SC~domains.
	
	The obtained, almost macroscopic spatial SDW--SC phase separation on a scale of $d\gtrsim 1$~\textmu m implies a rather weak influence of the SDW quantum critical point on  SC coupling. 
	Indeed, while in cuprate high-$T_c$ superconductors~\cite{InhBISCCO,InhBISCO2009,KresinReview2006,XRayNatPhys2012,InhHgBaCuO,Chubukov2015} and in transition metal dichalcogenides~\cite{Review1Gabovich,ReviewGabovich2002,MonceauAdvPhys} the SC--DW coexistence is more ''microscopic'' and~the corresponding $T_\text{cSC}$ enhancement is several-fold, in~organic superconductors, the $T_\text{cSC}$ enhancement by quantum criticality is rather weak, at $\sim 10$\%. 
	Note that in iron-based high-$T_c$ superconductors~\cite{ReviewFePnictidesAbrahams, ReviewFePnictides2}, e.g.,~in FeSe, the~$T_\text{cSC}$ enhancement by quantum criticality is also rather weak, at $\sim 10$\%. 
	A comparison of the observed~\cite{Mogilyuk2019,Grigoriev2023FeSe} $T_\text{cSC}$ anisotropy in thin FeSe mesa structures of various thicknesses with the numerical calculations of percolation anisotropy in finite-sized samples~\cite{Grigoriev2023FeSe}, similar to that in Section~\ref{sec:sizePerc}, suggests that the SC domain size in FeSe is also rather large, at $d\sim 0.1$~\textmu m, close to the nematic domain width in this compound. 
	Hence, similar to organic superconductors, in~FeSe and other iron-based high-$T_c$ superconductors, the large size of SC domains reduces the SC enhancement by critical fluctuations. 
	This observation may give a hint about raising the transition temperature in high-$T_c$ superconductors, which are always spatially inhomogeneous. 
	The knowledge of the parameters of SC domains also helps to estimate and even propose possible methods to increase the upper critical field and critical current in such heterogeneous superconductors, considered as a network of SC nanoclusters linked by Josephson junctions~\cite{KresinReview2006,Kresin2021}.
	
	
	To summarize, we have shown that the scenario in~which the first-order phase transition results in the spatial phase separation of SC and SDW in organic superconductors is self-consistent and also agrees with the available experimental data. 
	We estimated the size of SC domains $d$ by two different methods. 
	This estimate of $d>1$~\textmu m is consistent with various transport measurements, including the anisotropic zero resistance onset in thin samples~\cite{Kang2010,ChaikinPRL2014,Gerasimenko2014} and with angular magnetoresistance oscillations and magnetic-field-induced spin-density waves~\cite{ChaikinPRL2014,Gerasimenko2013}.
	We also discuss the relevance of our results, obtained for organic superconductors, to~high-$T_c$ superconductors, and~why the knowledge of SC domain parameters is important for increasing the transition temperature, the~critical magnetic field $H_{c2}$ and the critical current density in various heterogeneous~superconductors.  
	
	\vspace{6pt} 
	

	\authorcontributions{Conceptualization, P.D.G.; methodology, P.D.G. and V.D.K.; software, V.D.K.; validation, V.D.K. and S.S.S.; formal analysis, V.D.K.; investigation, V.D.K. and P.D.G.; writing---original draft preparation, V.D.K.; writing---review and editing, P.D.G. and S.S.S.; supervision, P.D.G. All authors have read and agreed to the published version of the~manuscript.}
	
	\funding{V.D.K. acknowledges the Foundation for the Advancement of Theoretical Physics and Mathematics ''Basis'' for grant \# 22-1-1-24-1, and~the RFBR grant \# 21-52-12027. The~work of S.S.S. was supported by the NUST "MISIS" grant no. K2-2022-025 in the framework of the federal academic leadership program Priority 2030. P.D.G.  acknowledges the State assignment \# 0033-2019-0001 and the RFBR grant \# 21-52-12043. }
	
	\institutionalreview{Not applicable.}
	
	\informedconsent{Not applicable.}
	%
	
	\dataavailability{Data will be provided on request.}
	
	
	\conflictsofinterest{The authors declare no conflicts of~interest.} 
	

		
	
	\appendixtitles{yes} 
	\appendixstart
	\appendix
	
	\section{Mean-Field Theory for~DW}
	
	\label{sec:AppendixA}
	
	The electronic Hamiltonian consists of the free-electron part $H_{0}$ and the
	interaction part $H_{\text{int}}$:
	\begin{equation}
		\begin{split}
			& H=H_{0}+H_{\text{int}}, \\
			& H_{0}=\sum_{\bm{k}}\varepsilon (\bm{k})a_{\bm{k}%
			}^{\dagger }a_{\bm{k}}, \\
			& H_{\text{int}}\ =\frac{1}{2}\sum_{\bm{k}\bm{k}^{\prime }%
				\bm{Q}}V_{\bm{Q}}a_{\bm{k}+\bm{Q}}^{\dagger
			}a_{\bm{k}}a_{\bm{k}^{\prime }-\bm{Q}}^{\dagger }a_{%
				\bm{k}^{\prime }}.
		\end{split}
		\label{H}
	\end{equation}%
	
	We consider the interactions at the wave vector $\bm{Q}$ close to
	the nesting vector $\bm{Q}_{0}=(\pm 2k_{F},\pi /b)$. If~the
	deviations from $\bm{Q}_{0}$ are small, we can approximate the
	interaction function as $V(\bm{Q})\approx V(\bm{Q}_{0})=U$.
	In the case of CDW, $U$ is the charge coupling constant, while for a SDW, $U$
	denotes the spin coupling constant. Next, in~the mean-field approximation, we
	introduce the order parameter
	\begin{equation}
		\begin{split}
			& \Delta _{\bm{Q}}=2U\sum_{\bm{k}}g(\bm{k}-%
			\bm{Q},\bm{k},-0), \\
			& g(\bm{k},\bm{k}^{\prime },\tau -\tau ^{\prime })=\langle
			T_{\tau }a_{\bm{k}^{\prime }}^{\dagger }(\tau ^{\prime })a_{%
				\bm{k}}(\tau )\rangle .
		\end{split}
		\label{eq:Delta_def}
	\end{equation}%
	
	Then, the final mean-field Hamiltonian, which we will study further, is
	\begin{equation}
		H_{\text{int}}\ =\sum_{\bm{k}\bm{Q}}\Delta _{\bm{Q}%
		}a_{\bm{k}+\bm{Q}}^{\dagger }a_{\bm{k}}+\operatorname{H.c.}+\operatorname{const}.  \label{Hmf}
	\end{equation}%
	
	The factor of $2$ in Equations~(\ref{eq:Delta_def}) and (\ref{nQ}) comes from the
	summation over two spin components. The~operators $a_{\bm{k}+%
		\bm{Q}}^{\dagger }$ and $a_{\bm{k}}$ correspond to the same
	spin component for a CDW, and~to different spin components for an SDW. From~Equations~(\ref{H}) and (\ref{Hmf}), using the standard equation for the operator
	evolution, $i\hbar \diff{\hat{A}}/\diff{t}=\left[ \hat{A},\hat{H}\right] $, one obtains
	the equations of motion for the Fourier transform of the Green's function $%
	g(\bm{k},\bm{k}^{\prime },\tau -\tau ^{\prime })=\int \diff \omega /(2\pi)
	e^{i\omega \left( \tau -\tau ^{\prime }\right) }g(\bm{k}^{\prime },%
	\bm{k},\omega)$:%
	\begin{equation}
		\lbrack i\omega -\varepsilon (\bm{k})]g(\bm{k}^{\prime },%
		\bm{k},\omega )-\sum_{\bm{Q}}\Delta _{\bm{Q}}g(%
		\bm{k}^{\prime },\bm{k},\omega )=\delta _{\bm{k}%
			^{\prime },\bm{k}}.  \label{SCGF}
	\end{equation}
	
	In the metallic phase, $\Delta _{\bm{Q}}$, given by Equation (\ref%
	{eq:Delta_def}), vanishes after the thermodynamic averaging denoted by
	triangular brackets in Equation~(\ref{eq:Delta_def}). If~the DW at wave vector $%
	\bm{Q}_{0}$ is formed, the~order parameter $\Delta _{\bm{Q}%
	}\neq 0$ for $\bm{Q=Q}_{0}$, while for $\bm{Q\neq Q}_{0}$,\
	the average $\Delta _{\bm{Q}}=0$. The~spatial variation in the order
	parameter $\Delta \left( \bm{r}\right) =\int \Diff{3}{\bm{q}}/(2\pi)
	\Delta \left( \bm{q}\right) e^{i\bm{q}\cdot\bm{r}}$ is described by the deviation $\bm{q=Q-Q}_{0}$ of the
	DW wave vector $\bm{Q}$ from its value $\bm{Q}_{0}$,
	corresponding to the maximum  susceptibility. If~we now set $\Delta _{%
		\mathit{Q}_{1}}=\Delta \delta _{\mathit{Q}_{1},\pm \bm{Q}}$, where $%
	\bm{Q}=\bm{Q}_{0}+\bm{q}$ with $|\bm{q}|\ll
	k_{F}$, the~equations of motion (\ref{SCGF}) can be solved, giving
	\begin{equation}
		g(\bm{k}-\bm{Q},\bm{k},\omega )=-\frac{\Delta _{%
				\bm{Q}}}{(\omega +i\varepsilon _{+})^{2}+\varepsilon
			_{-}^{2}+|\Delta _{\bm{Q}}|^{2}},  \label{g}
	\end{equation}%
	where $\varepsilon _{\pm }=\varepsilon ^{\pm }({\bm{k}},{\bm{%
			k}}-{\bm{Q}})$ are given by Equation~(\ref{epspm}). From~Equations (\ref%
	{eq:Delta_def}) and (\ref{g}), we obtain the self-consistency Equation (\ref%
	{Delta_SC}), omitting the subscript $\bm{Q}$ in $\Delta_{\bm{Q}}$:
	\begin{equation}
		\Delta =-TU\sum_{\bm{k}\omega }\frac{\Delta }{(\omega +i\varepsilon
			_{+})^{2}+\varepsilon _{-}^{2}+|\Delta |^{2}}%
		.
	\end{equation}
	
	The mean-field Hamiltonian given by Equations~(\ref{H}) and (\ref{Hmf}) decouples
	to a sum over $\bm{k}$ of $2\times 2$ matrices. Their
	diagonalization gives the new quasiparticle dispersion given by Equation (\ref%
	{EnSp}). Hence, the~order parameter $\Delta _{\bm{Q}}$ defined in
	Equation~(\ref{eq:Delta_def}) has the physical meaning of the DW energy gap for
	the case of perfect nesting. One could define the order parameter in a
	different way as
	\begin{equation}
		n_{\bm{Q}}=\frac{\Delta _{\bm{Q}}}{U}=2\sum_{\bm{k}%
		}\langle T_{\tau }a_{\bm{k}^{\prime }}^{\dagger }(\tau ^{\prime })a_{%
			\bm{k}}(\tau )\rangle ,  \label{nQ}
	\end{equation}%
	which has the physical meaning of electron density $n_{\bm{Q}}$ at
	wave vector $\bm{Q}$. The~latter couples to the external potential $%
	V_{\bm{Q}}$ at the same wave vector in the Hamiltonian: $\delta
	H=\delta F=-\sum_{\bm{Q}}n_{\bm{Q}}V_{\bm{Q}}$. The~equilibrium value of the DW order parameter $\Delta _{\bm{Q}}=Un_{%
		\bm{Q}}$ in the presence of an external field $V_{\bm{Q}}$\ can
	be obtained from the minimization of the total free energy $F_{tot}=F+\delta
	F$, where the free energy $F$ without an external field is given by Equation~(\ref{F}%
	) at $\Delta _{\bm{Q}}\rightarrow 0$:
	\begin{equation}
		\frac{\partial F_{tot}}{\partial n_{\bm{Q}}}=-V_{\bm{Q}}+U%
		\frac{\partial F}{\partial \Delta _{\bm{Q}}}=0,  \label{F1}
	\end{equation}%
	or%
	\begin{equation}
		-V_{\bm{Q}}+U^{2}n_{\bm{Q}}\left[ A\left( T,{\bm{Q}}%
		\right) +B\left\vert \Delta _{\bm{Q}}\right\vert ^{2}+\dots \right] =0.
		\label{F2}
	\end{equation}%
	
	Hence, the~electronic susceptibility just above the DW phase transition
	temperature $T_\text{cDW}$, where $\Delta _{\bm{Q}}=0$, is related to the
	coefficient $A\left( T,{\bm{Q}}\right) >0$ of the Landau--Ginzburg expansion:
	\begin{equation}
		\chi \left( \bm{Q}\right) =\frac{n_{\bm{Q}}}{V_{\bm{Q%
		}}}=\frac{1}{A\left( T,{\bm{Q}}\right) U^{2}}.  \label{chi1}
	\end{equation}%
	
	At the DW transition temperature $T=T_\text{cDW}$, the coefficient $A\left( T,{%
		\bm{Q}}\right) =0$ for some ${\bm{Q}}$. Hence, the~DW wave
	vector $\bm{Q}$ corresponds to the minimum of $A\left( T_\text{cDW},{%
		\bm{Q}}\right) $ or to the maximum of susceptibility $\chi \left( 
	\bm{Q}\right) $ in Equation~(\ref{chi1}). Near~this extremum, one can
	expand Equation~(\ref{chi1}) over the deviation $\bm{q=Q-Q}_{0}$ of the
	DW wave vector $\bm{Q}$ from its optimal value $\bm{Q}_{0}$:
	\begin{equation}
		\chi \left( \bm{Q}\right) =\frac{\chi \left( \bm{Q}%
			_{0}\right) }{1+\xi ^{2}\bm{q}^{2}},  \label{chiQ}
	\end{equation}%
	which gives the estimate of the DW coherence length $\xi $.
	
	Below the phase transition temperature $T_\text{cDW}$, Equation~(\ref{F2}) gives
	\begin{equation}
		\chi ^{-1}\left( \bm{Q}\right) =U^{2}\left[ A\left( T,{\bm{Q}%
		}\right) +B\left\vert \Delta \right\vert ^{2}+C\left\vert \Delta \right\vert
		^{4}+..\right] \to \infty ,  \label{chi2}
	\end{equation}%
	which corresponds to a finite $\Delta _{\bm{Q}}$ at vanishing $V_{%
		\bm{Q}}$. Nevertheless, one can find the differential susceptibility
	\begin{equation}
		\chi ^{-1}\left( \bm{Q}\right) =\frac{\diff{V_{\bm{Q}}}}{\diff{n_{\bm{Q}}}}
		=\frac{\partial ^{2}F}{\partial n_{\bm{Q}}^{2}}=U^{2}%
		\frac{\partial ^{2}F}{\partial \Delta _{\bm{Q}}^{2}},  \label{chid}
	\end{equation}%
	which generalizes Equation~(\ref{chi1}).


\reftitle{References}




\end{document}